\documentclass[twocolumn,showpacs,preprintnumbers,amsmath,amssymb]{revtex4}
\usepackage{graphicx}
\usepackage{dcolumn}
\usepackage{bm}
\begin{document}
\title{A Simulation on Vertically Shaken Granular Layers}
\author{Y.~K.~GOH}
\author{R.~L.~JACOBS}
%
%
\affiliation{Department of Mathematics,\\
           Imperial College,
           London SW7~2AZ.}
\date{\today}
%
%
\begin{abstract}
A hybrid model of molecular dynamics and continuum mechanics is introduced
 to study a system of vertically shaken granular layers.
Despite the simplicity the model shows pattern formation in the
 granular layers due to the formation of heaplets.
We show from a simple analysis that the onset of pattern formation is
 density dependent and 
 this result is justified by the subsequent computer simulations.
Our simulations also show that the heaping process
 can be divided into three stages: an
 early stationary stage, an intermediate growing stage
 and a late-time saturated stage.
In the early stage, the average volume of the heaplets remains almost unchanged
 until the system crosses over to the intermediate growing stage.
The length of time that system remains in the early stage defines the 
 onset time of the instabilities, $k_0$ which depends on the shaking 
 intensity, $\gamma$.
In the growing stage, 
 the average volume of the heaplets grows with 
 time and can be approximated  by a power law
 with a shaking intensity dependent growth exponent, $z$.
From our simulation, we show that the growth exponent, $z$ depends 
 linearly on the shaking intensity, $\gamma$.
There is a critical shaking intensity  $\gamma_c \simeq 14$, such that
 for shaking intensities greater than $\gamma_c$, $z$ cannot be properly
 defined.
The late-time saturated stage is where most of the particles are
 trapped in a big heap and this big heap is in equilibrium with the
 surrounding granular gas.
\end{abstract}
\pacs{61.43.Gt,45.70.Qj,83.80.Fg} 
\maketitle 
\section{Introduction}\label{sec:intro}
We are interested in pattern formation in a thin layer of granular 
 particles upon a plate.
We begin by distinguishing between experiments on systems in which 
 the plate is vertically vibrated~\cite{Melo:1994Jan}
 or in which the plate is tapped~\cite{Duran:2001Dec} or
 shaken.
In the vertically vibrating systems, the particles dance upon the oscillating
 plate and the surface of the layer becomes patterned with
 regions characterised by different modes of oscillation or by the
 the same modes of oscillation but different phases.
The features described as oscillons~\cite{Umbanhowar:1997Oct}
 and the interface between regions~\cite{Blair:2000May} are 
 both of interest here.
In the tapped or shaken system there is a long period between taps/shakes
 in which the plate is at rest and the particle layer relaxes in 
 this period to a static configuration in which the surface of the layer
 is marked by ripples or heaplets.
In other words the perturbations are discrete and are followed by periods of
 stasis in distinction to the vibrated
 system which is continuously moving.

Following the original observation of square and striped patterns
 by Melo {\em et. al}~\cite{Melo:1994Jan,Melo:1995Nov} in the vibrating system,
 there  have been many experimental
 observations~\cite{Umbanhowar:1997Oct,Umbanhowar:1998xxx,Aranson:1999Jan}
 and theoretical discussions~\cite{Tsimring:1997Jul,Shinbrot:1997Oct,Aranson:1998xxx,Aranson:1999Feb,Blair:2000May} for this type of  pattern formation.
All these works and many others have contributed greatly and
 also establish a good understanding of the dynamics of the
 pattern formation of granular layer under continuous vibration.
On the other hand in the tapped/shaken system,
 Duran has observed the formation of ripples~\cite{Duran:2000May}
 and granular heaplets~\cite{Duran:2001Dec} in a tapped granular layer.
However, there is less literature available on
 this class of experiments, and  our paper is an attempt to fill this gap.

We now further distinguish between vertically shaken granular
 layers and tapped granular layers.
In the shaken system the plate moves up and then is withdrawn sharply
 to a neutral position.
When the plate reaches the highest point, the particles
 continue their motion and 
 then travel ballistically, falling back to the stationary plate,
 rebounding and then settling
 down and relaxing into a static configuration.
The plate is subsequently moved up again and the process repeated.
The lateral redistribution, which produces the patterns, takes place when 
 the falling particles rebound from the plate and relax.
On the other hand
 in the tapped system the plate scarcely moves but, as in the Newton's cradle
 experiment, the tap is imparted
 impulsively to the particles which jerk off the plate, fall back,
 rebound and relax.
Most of the lateral redistribution takes place at the jerk. 
However, shaken and tapped granular layers also share some common properties
 such as the formation of isolated granular heaplets, which coarsen 
 with the number of taps/shakes, and the rate of coarsening of the heaplets
 increases as the intensity of the shocks increases.
We have discussed tapped layers in a previous article~\cite{Goh:2002Oct}.
The present is concerned mainly with shaken granular layers.

This article is organised as follows:
 first, we introduce a simple model to study shaken granular layers.
The model is based on a hybrid of a molecular dynamics description
 and continuum modelling of a granular sheet.
Then we show typical patterns produced by the model and these patterns 
 resemble typical experimental morphologies
 of shaken granular layers with isolated heaplets.
An analysis is performed and leads to the prediction that the
  depinning shaking intensity of the granular layers is density dependent,
  which is later justified from the simulations.
Later we show from the simulation that the coarsening of the pattern
  comes in three stages:
  an early stationary stage, a growing 
  stage and a saturated stage.
In the early stage, the layer remains flat, with small fluctuation 
 superimposed, for a short duration $k_0$ which depends
 on the shaking intensity $\gamma$.
The early stage is characterised by a small value $V$ (defined later),
 which measure the degree of heaping.
Then we discuss the cross-over time from the early stage to the growing stage,
 which indicates the onset time $k_0$ of instabilities.
In the growing stage, we approximate the growth of the average volume
 of heaplets by a power law.
The growth exponent $z$ is $\gamma$-dependent and the power law approximation
 fails when $\gamma$ is larger than a critical value, $\gamma_c$.
Finally, the late-time saturated stage is where all the granular heaplets
 are merged into a big heap and the big heap is in equilibrium with
 a scattering of particles on the plate which can be described as a 
 granular gas.
The time of onset of saturation is dependent on the size of
 the system.
In our simulations, the system size is small (with periodic
 boundary conditions),  and as a result the time of onset will almost
 certainly be small compared to that experimentally observed in real systems.

\section{Model}\label{sec:model}
The spatial distribution of a granular layer can be described by a continuum 
 field variable $n(\vec{r},t)$, where $\vec{r}=(x,y)$ are the 
 Cartesian coordinates parallel to the plate.
The number density $n(\vec{r},t)$ also represents the layer thickness if 
 we take the average diameter of grains as unity
 assuming no compactification throughout the whole shaking
 process~\cite{Goh:2002Oct}.
When the system is shaken the layer moves off the plate to
 height $h(\vec{r},t)$ measured from the {\em stationary} neutral
 position of the plate.
There is also an associated velocity variable
 $u(\vec{r},t) = \partial h(\vec{r},t)/\partial t$
 and $u(\vec{r},t)$ then
 represents the average vertical velocity of the grains at the point
 $\vec{r}$.
As in Rothman~\cite{Rothman:1998Feb} the velocity field
 at each point is coupled dissipatively with the field at
 neighbouring points through a Laplacian,
 and the dynamics can be described by a generalisation of Newton's
 second law,
 \begin{equation}\label{eq:u}
  \frac{\partial u}{\partial t}=\nu \nabla^2 u - g + B(h,u,\alpha).
 \end{equation}
Here $\nu$ is a dissipative coefficient; 
 it represents the lateral exchange of velocities when particles collide 
 with each other and $g$ is the acceleration due to gravity.
The function $B$ specifies
 how the layer bounces when it hits the plate.
The bouncing term $B$ is zero except when $h=0$
 and it is defined by the mapping
 \begin{equation}\label{eq:B_map}
   u \rightarrow -\alpha u,
 \end{equation}
 where $\alpha$ is the coefficient of restitution.
The same equations here have been used by Rothman\,\cite{Rothman:1998Feb}
 and Moon~{\em et al.}\,\cite{Moon:2002Jun} to describe the phenomena of 
 oscillons and stripes structures in  a vertically vibrated sand layer.

The coefficient of restitution $\alpha$ is expected to depend on $n$
 because the energy dissipation is greater in regions of higher local density.
However, the precise form of $\alpha(n)$ is not known, 
 except that it should be a monotonic decreasing function of $n$.
In fact, which we will show in the following section, it is 
 possible to deduce roughly the functional form of $\alpha(n)$ 
 from experiments by measuring the depinning shaking intensities
 for homogeneous granular layer with different thickness.
In this paper, for the sake of definiteness,
 $\alpha(n)$ is chosen to be a Lorentzian, {\em i.e.}
 \begin{equation}\label{eq:alpha}
   \alpha(n) = \frac{n_0^2}{n^2 + n_1^2},
 \end{equation}
 where $n_0$ and $n_1$ are parameters.

In the shake the plate moves up and is then withdrawn sharply.
The particles continue in free flight, all with the same velocity, and 
 fall under gravity until they hit the plate at rest, all with
 the same velocity $-U$.
They then rebound with {\em local} velocity $u(\vec{r},t) = \alpha(n)U$
 which depends on the {\em local} density $n(\vec{r},t)$ because of the
 density dependence of the coefficient of restitution $\alpha$.
It is useful to use $\gamma=U/U_0$ as a dimensionless measure
 of the shaking intensity, here $U_0 = 3\,\mathrm{cm/s}$ being a very crude
 estimate of a typical take-off velocity of the grains
 when the plate is withdrawn.
The purpose of $U_0$ is to allow us to present the data on a reasonable scale.

After the shake, the particles undergo motion as given by Eq.(\ref{eq:u}).
When the particles at a point fall back to the plate again, 
 they bounce and redistribute onto the neighbourhood.
To describe this, we discretise the system into $L\times L$ squares of
 side $a$ with
 the square at $\vec{r}_i$ containing $N_i = n(\vec{r}_i,t)\,a^2$ particles.
Then we implement the redistribution in the following way:
\begin{enumerate}
\item{} We define an integer $N_i$ equal to the the next integer greater than
        $n(\vec{r}_i,t)$, which is the number of particles
        on the rebounding site.
\item{} We surround the rebounding site $\vec{r}_i$ by a disc of
        radius $R_{max}$~\cite{Goh:2002Oct}, which is given by
        \begin{equation}\label{eq:Rmax}
           R_{max}=\xi \left(\frac{u(\vec{r}_i,t)}{U_0}\right)^2,
        \end{equation} 
        where $\xi$ is a parameter which controls the distance travelled by the 
        particles.
\item{} We specify a hopping angle $\phi$ by choosing a random number
        in the range $[0,2\pi]$ with equal probability.
\item{} We specify a hopping range $R$ by choosing a random number 
        in the range $[0,R_{max}]$  with equal probability.
\item{} The angle $\phi$ and hopping range $R$ together define
        a receiving site $\vec{r'}_j$ such that $|\vec{r'}_j-\vec{r}_i|=R$
        and the angle between the vector $\vec{r'}_j-\vec{r}_i$ and a
        fixed reference direction is $\phi$.
\item{} The process is repeated $N_i$ times but receiving sites may 
        be repeated and the rebounding site itself may be a receiving site.
\item{} On each repetition (except the last) the density $n(\vec{r}_i,t)$ on
        the rebounding site is reduced by unity and the density on the 
        receiving site is increased by unity. On the last repetition, 
        leftovers are redistributed.
\end{enumerate}
The redistribution rules can be summarised by first defining $N_i$ and the
 size of the leftovers, {\em i.e.}
 ${\Delta n_i \equiv n(\vec{r}_i,t)-N_i+1}$,
then repeat the following $N_i$ times.
 \begin{equation}\label{eq:redistribution}
   \left.
     \begin{array}{lll}
       n(\vec{r}_i,t)
     & \rightarrow
     & n(\vec{r}_i,t) - \Delta_j
    \\ n(\vec{r'}_j,t)
     & \rightarrow
     & n(\vec{r'}_j,t) + \Delta_j,    
     \end{array} \right\}
  \qquad \mathrm{for}\, j=1,2,\dots,N_i,
 \end{equation}
where 
 \begin{equation}
   \Delta_j= \left\{
      \begin{array}{ll}
        1 & \quad j=1,2,\dots,N_i-1\\
        \Delta n_j & \quad j = N_i.
      \end{array}\right.
 \end{equation}
The process conserves the number of particles in the redistribution
 as it should.
If $R_{max}$ is greater than system  size the physical reality is 
 not well-represented.
The mass transfer after redistribution can be describe by a continuity
 equation $\partial_t n = -\nabla \cdot \vec{J}$. 
The constitutive current equation is given by 
 \begin{equation}\label{eq:J}
   \vec{J} = \left\{
   \begin{array}{ll}
     -\eta(\beta |\nabla n|^2 - 1)\nabla n, 
     &
     |\nabla n| > \frac{1}{\sqrt{\beta}} 
     \\
     0, &
     |\nabla n| \leq \frac{1}{\sqrt{\beta}} ,
   \end{array}
   \right.
 \end{equation}
 where $1/\sqrt{\beta}=\tan\theta_c$ is the critical slope of the heaplet
 side and $\eta$ sets the diffusion rate.
Here the angle of repose $\theta_c$ is chosen to be $\pi/6$.
The constitutive equation is constructed in such a way that when the gradient 
 is greater than a critical slope, 
 mass current will flow down hill to smooth out density fluctuation,
 but no mass is transfered if the gradient is less than the critical slope,
 because this corresponds to a stable situation.

\section{Simulation}\label{sec:simulation}
The model is studied on a $L \times L$ square lattice with 
 periodic boundary condition.
Initially the granular layer is prepared with the density $n$
 randomly distributed about an average value $\bar{n}$.
At the beginning of each shake, each site is assigned a
 velocity $u(\vec{r},t)=\alpha(n(\vec{r},t)) \gamma U_0$.
We choose the parameters $n_0$ and $n_1$ in Eq.(\ref{eq:alpha}) for
 $\alpha(n)$ such that $n_0^2 =1/3$ and $n_1^2=2/3$.
The value of the lattice spacing $a$ is set equal to the 
 average diameter of the particles
 and the hopping parameter $\xi$ is set to
 $75\sqrt{3} U_0^2/2g$~\cite{Goh:2002Oct}.
The reason we choose $\xi$ differently from reference~\cite{Goh:2002Oct}
 and $U_0 = 3 \,\mathrm{cm/s}$ is  to allow us present data on a reasonable 
 scale.
Then Eq.(\ref{eq:u}) and the  equation of continuity with Eq.(\ref{eq:J})
 are solved using a Runge-Kutta method.
On each shaking cycle, these equations are iterated until no more lateral  mass
 transfer occurs at each site because of the rebound
 and because of the local gradient on each site is less
 than the tangent of the angle of repose, $\theta_c=\pi/6$ with 1\% cutoff.

\section{Results}\label{sec:results}
\subsection{Morphology}
\begin{figure}[!htb]
\vspace{0.5cm}
\resizebox{\hsize}{!}{%
  \includegraphics{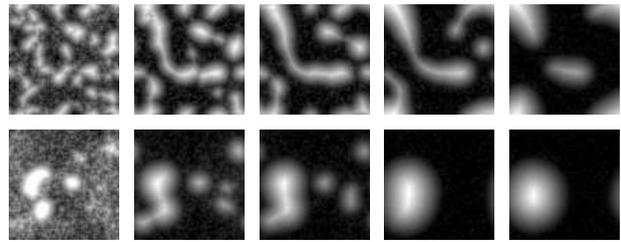}
}
\caption{Two simulations of a bouncing granular layer corresponding to
  different tapping intensities.
 The top panel correspond to a lower shaking intensity $\gamma = 4.0$,
  and the lower panel corresponds to a higher shaking intensity $\gamma=7.0$.
 Each column is taken at a specific time and from left to right the 
  corresponding number of shakes is $k=5,10,15,30$ and $50$.}
\label{fig:density}
\end{figure}

Fig.~\ref{fig:density} shows typical results of the pattern formation in
 two different simulations.
The top panel corresponds to a lower shaking intensity ($\gamma = 4.0$),
 and the lower panel corresponds to
 a higher shaking intensity ($\gamma = 7.0$).
Bright regions correspond to a higher density region of the granular heaplets
 in the granular layer.
Dark regions correspond to regions of low densities
 of freely moving interstitial 
 granular particles (granular gases).
Both panels show an effective surface tension as suggested in
 reference~\cite{Goh:2002Oct}.
The effective surface tension depends on the shaking intensity.
In the lower panel a greater effective surface tensions is associated 
 with a higher shaking intensity,
 and this greater surface tension produces more compact heaplets,
 which in agreement with the findings
 in reference~\cite{Goh:2002Oct}.

\subsection{Depinning Shaking Intensity}\label{sub:depin}
\begin{figure}[!htb]
\vspace{0.5cm}
\resizebox{0.9\hsize}{!}{%
  \includegraphics{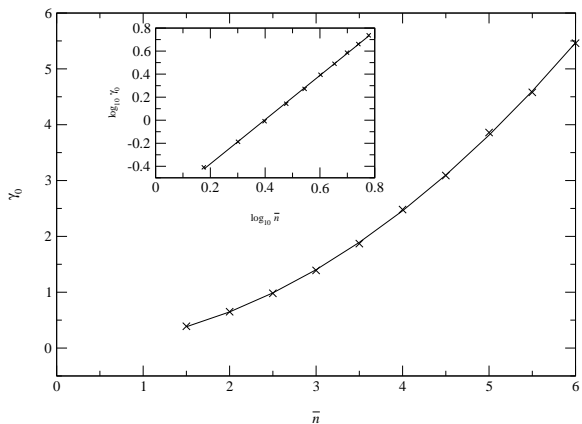}
}
\caption{Variation of $\gamma_0 $ 
 for different values of $\bar{n}$.
 The inset shows the log-log plot for $\gamma_0$ against $\bar{n}$.
 The linear regression fit line for the inset gives a slope of $1.91$.}
\label{fig:g_depin}
\end{figure}

In order to form heaplets in the granular layer, 
 the shake intensity  not only needs to be strong enough to lift
 grains off the plate but it must also be able to redistribute grains
 so the hopping range must greater than the lattice spacing $a$.
One can predict the shaking intensity $\gamma_0$ for 
 such a depinning process from  a mean field argument.
Let the thickness of the granular layer be homogeneous throughout 
 the system and be equal to $\bar{n}$. 
After the first shake, the rebound velocity of the layer is 
$\bar{u} = \alpha(\bar{n})\gamma U_0$ everywhere.
We can obtain 
 the depinning shaking intensity $\gamma_0$ from
 Eq.(\ref{eq:Rmax}) and setting $R_{max}(\bar{n}) = a$, so that
 \begin{equation}\label{eq:g_depin}
   \gamma_0= C (\bar{n}^2 + n_1^2)
 \end{equation}
 and 
 \begin{displaymath}
   C = \left(\frac{a}{\xi}\right)^{1/2} \frac{1}{n_0^2U_0}.
 \end{displaymath}
However, one might expect the relation given
 by Eq.(\ref{eq:g_depin}) to be not strictly
 quadratic in $\bar{n}$ when  fluctuations are important.
Fig.~\ref{fig:g_depin} shows the variation of $\gamma_0$ against
 average density $\bar{n}$ from the simulation.
The graph show that the relation between $\gamma_0$ and $\bar{n}$
 is in good agreement with the prediction.
From the inset of Fig.~\ref{fig:g_depin} shows that the exponent 
 is only slightly less than $2$ ($\sim 1.91$). 
This result is due to the 
 specific form for the effective coefficient of  restitution $\alpha(n)$
 we chose in Eq.(\ref{eq:alpha}).
However, this also suggests a possibility that one may construct
 the functional form of $\alpha(n)$ from experiments.
This can be done by noting the relation between
 depinning shaking density $\gamma_0$, which can be measured from experiments,
 and the corresponding average density of layer $\bar{n}$ is given by
 $\gamma_0=(a/\xi)^{1/2} \, \left[\alpha(\bar{n})\right]^{-1}$.
We present this as a challenge to experimentalists.

\subsection{Average volume of heaplets}\label{sub:V}
\begin{figure}[!htb]
\vspace{0.5cm}
 \resizebox{0.9\hsize}{!}{%
   \includegraphics{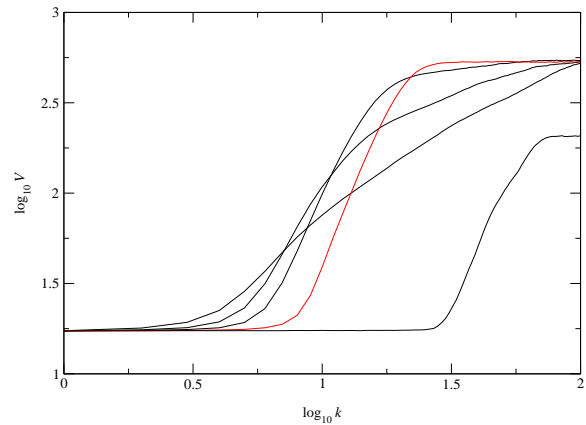}
 }
 \caption{Log-Log plot of heaplet size, $V$ against number of shakes, $k$ 
  for different values of $\gamma$. 
  From left to right the corresponding value of $\gamma$ for each
   curve is 6, 8, 10, 12 and 14.
  Each curve is averaged over 10 different runs.
  Note the break-down of power-law growth
  in the intermediate region for the curves with $\gamma > 14$.}
\label{fig:loglog}  
\end{figure}

We are also interested in how the size of heaplets grows in time.
The relevant length scale can be obtained as the average of the thickness 
 at site $i$  weighted by its local volume.
We know that the thickness is $n_i$ 
 and the local volume element of site $i$ is $V_i=n_i\sigma$, 
 where site area $\sigma$ is the square of the lattice constant,
 then
 \begin{equation}\label{eq:l}
 l = \frac{\sum_i h_i\, V_i}{\sum_i V_i} = \frac{\sum_i n_i^2}{\sum n_i}.
 \end{equation}
Hence the volume of the heaplet is $V=l^3$ 
and Fig.~\ref{fig:g2g1} and Fig.~\ref{fig:loglog} shows the typical linear
 and log-log plots 
 of heaplet volume $V$ for different values of $\gamma$ 
 as a function of the number of shakes, $k$.

In each plot in Fig.~\ref{fig:loglog} there are three different stages 
 and each stage corresponds to a different underlying process.
In the early stage, the volume remains almost unchanged in time.
In this stage, the fluctuations in the 
 randomly distributed granular layer are not strong enough
 to trigger coalescence.
As a result, the layer remains nearly flat and has a small value of $V$.
The second stage is the heaplet formation stage, which corresponds to
 the growing regions in the plots.
In this stage, high density regions of the granular layer 
 act as a kinetic energy sinks,
 many random hopping granular particles hop into these sinks 
 and are trapped there.
As a result of capturing particles, these high density regions grow
 and merge to form bigger heaplets.
Finally, the late time stage is where all the the heaplets have joined 
 into a single big heap and stop growing.
There are some remaining individual particles (granular gas)
 hopping randomly about 
 and occasionally encountering the big heap and getting trapped there.
Occasionally particles on the big heap will leave the heap and join the 
 surrounding granular gas.
When the number of particles leaving the big heap is balanced by those
 entering the big heap, the system is now in an equilibrium situation
 which very similar to that of droplets in a supersaturated
 vapour.
However there are very few individual particles involved in the
 motion across boundaries during equilibrium and 
 therefore the fluctuations of the volume of big heap are small as observed
 in Fig.~\ref{fig:loglog} and Fig.~\ref{fig:g2g1}.
Of course the size of the saturated heap depends on the size of the
 simulated system and the average number density of particles.
Larger system sizes will result in a later saturation and bigger saturated 
 heaps.

\subsection{Growth Exponents}\label{sub:growth}
\begin{figure}[!htb]
\vspace{0.5cm}
 \resizebox{0.9\hsize}{!}{%
   \includegraphics{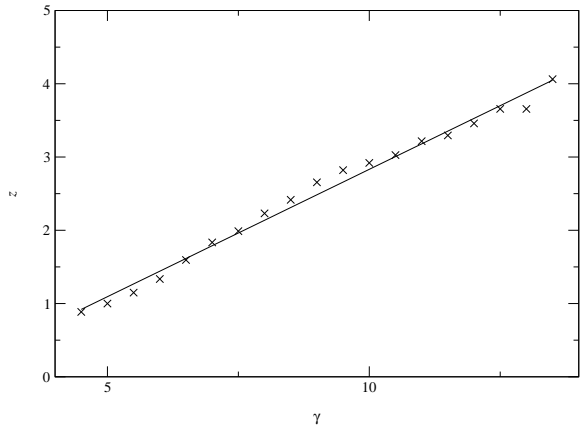}
 }
 \caption{The growth exponent $z$ as a function of  shaking intensity $\gamma$.
 The straight line is the linear regression fit for the data, 
 the slope of straight line is 0.348.}
\label{fig:g_z}  
\end{figure}

In the growing stage, the growth exponent $z$ is $\gamma$-dependent.
Fig.(\ref{fig:g_z}) shows the growth exponent $z$ for different
 values of $\gamma$.
As we can see there is a trend of increasing growth exponent as 
 the shaking intensity $\gamma$ is increased.
The relation between the growth exponent and the shaking intensity is 
 roughly linear and is given by $z \simeq 0.348\gamma - 0.644$, 
 as shown as the straight line in Fig.(\ref{fig:g_z}).
However, the relation breaks down when $\gamma \gtrsim 14$.
The reason of the break-down of the power law approximation will
 be discussed in the following section.
We can understand the dependence of $z$ to $\gamma$ from the fact that
 larger shaking intensities give larger kinetic energy to allow the 
 system explore greater regions of phase space.
As a result, systems with larger shaking intensities are able to find
 stable configurations easier than systems with lower shaking intensities.
Of course, the final stable configuration is a single heap which favour
 greater dissipation.

\begin{figure}[!htb]
\vspace{0.5cm}
\resizebox{0.9\hsize}{!}{%
  \includegraphics{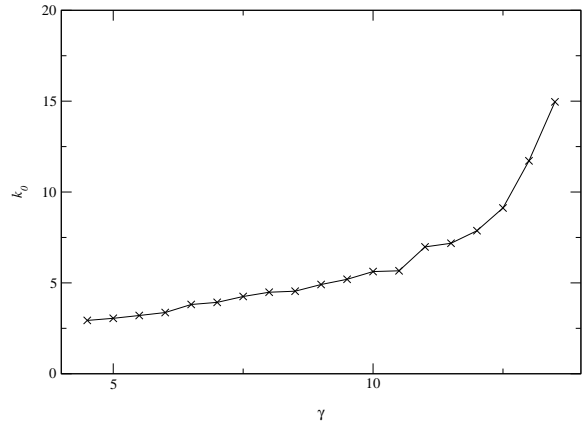}
}
\caption{Variation of $\log_{10} k_0$ against $\log_{10} \gamma$.
 It is obvious that $k_0$ increases with $\gamma$.}
\label{fig:g_k0}  
\end{figure}

\subsection{Onset Time}\label{sub:t_onset}
\begin{figure}[!htb]
\vspace{0.5cm}
\resizebox{0.9\hsize}{!}
 { \includegraphics{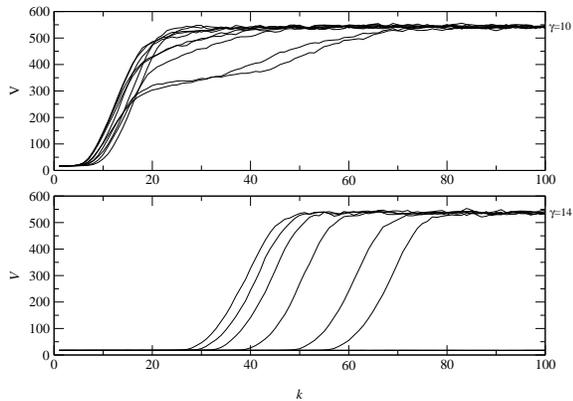} }
\caption{
  Each panel shows the time series of $V$ for several different runs.
  The top panel corresponds to $\gamma = 10$.
  For shaking intensities $\gamma < 14$, the onset time of the 
    instabilities is consistent for different initial configurations.
  The irregular slow-down of the growth in some of the curves is due to 
    the formation of more than one heaplet.
  The merging process of the heaplets is slow, this is due to the strong 
    dissipation in each heaplet and particles in heaplets
    are less mobile than those not in heaplets.
  The lower panel corresponds to a shaking intensity
  $\gamma$ where the scaling law
    fails ($\gamma = 14$).
  At shaking intensities $\gamma \ge 14$, one loses the predictability of 
    the onset time of the instabilities.
  The early stage where the granular layer is relatively flat is metastable
    and the system can be trapped in this stage and then suddenly jump
    at an unpredictable time to the saturated stage within a few shakes.
  The time of the jump appears to depend sensitively on the initial conditions.
 }
\label{fig:g2g1}
\end{figure}

We also note from Fig.~\ref{fig:loglog} that the onset of the growing
 stage occurs at different times for different $\gamma$'s.
Generally the onset time $k_0$ increases as $\gamma$ increases.
Fig.(\ref{fig:g_k0}) plots $k_0$ against $\gamma$ and shows that $k_0$ is
 an increasing function of $\gamma$.
However, for $\gamma$ greater than a critical value $\gamma \gtrsim 14$,
 heaplets can start forming at any time  hence no proper $k_0$ can be defined.
The same reason that $k_0$ loses its predictability also
 explains why the power law approximation fails
 for $\gamma > \gamma_c$.
This can be clearly seen from Fig.\ref{fig:g2g1}.
Each panel of Fig.~\ref{fig:g2g1} contains time series of $V$ for 10
 different runs where each run has a different initial configuration
 but the  same shaking intensity.
The top panel corresponds to the case where $\gamma < \gamma_c$
 $(\gamma = 10)$.
It is clear that there is a well defined onset time of instabilities at 
 $k\simeq 6$ and consistent growth after the instability starts.
The irregular slow-down of the growth
 during the cross-over from growing stage to saturated 
 stage is due to the formation of multiple metastable heaps
 and the merging process of these heaps is slow.
On the other hand, curves in the lower panel do not have a well defined onset
 time, instabilities in each curve appear at a different time.
In fact, some curves remain trapped in the metastable early stage
 and instabilities never appear through out the simulations.

%

\section{Conclusion}
We have introduced a simple model to study vertical shaken granular layers.
The model shows the shaking intensity for the depinning
 of granular layer is related to the layer thickness and 
 to the effective coefficient of restitution of the granular layer.
The dynamics of heaplet formation in shaken granular layer can
 be divided into three stages:
 an early stationary stage ($\gamma$-independent),
 an intermediate growing stage ($\gamma$-dependent), and
 a saturated/equilibrium stage.
In the growing stage, the growth exponent $z$ is an increasing 
 function of $\gamma$.
The onset time $k_0$ increases as $\gamma$ increases up to a
 critical value of $\gamma_c (\simeq 14)$ where
 heaplet growth can start at any time dependent on details of 
 the initial configuration.

\bibliographystyle{h-physre}
\bibliography{physics}

\end{document}